\documentclass[12pt]{article}
\pdfoutput=1
\usepackage{amsmath}
\usepackage{ifxetex,ifluatex}
\usepackage{graphicx,psfrag,epsf,grffile}
\usepackage{enumerate}
\usepackage{natbib}
\usepackage{url} 
\usepackage{multirow}

\newcommand{\blind}{0}
\newcommand{\supplement}{0}
\graphicspath{ {figures/} }
\usepackage{floatpag}
\usepackage{amssymb}
\begin{document}

\def\spacingset#1{\renewcommand{\baselinestretch}%
{#1}\small\normalsize} \spacingset{1}


\if1\supplement
{
 \vspace{3mm}
  \begin{center}
  \LARGE\bf Supplementary Material for \\
  \medskip
    {\LARGE\bf Downstream Effects of Upstream Causes}    
\end{center}
  \medskip
}
\fi

\if0\supplement{
\if0\blind
{
  \title{\bf Downstream Effects of Upstream Causes}
 
\author{
        Bradley C. Saul \thanks{\noindent We would like to thank Dr.~Rebecca Benner and the Nature
Conservancy of North Carolina for compiling and providing the data. We
would also like to thank all those who collected and collated these data
in order to protect the Cape Fear watershed. The causal inference with
interference research group at
\if0\blind{UNC (Brian Barkley, Sujatro Chakladar, and Wen Wei Loh) plus Mary Kirk Wilkinson }\fi \if1\blind{BLINDED }\fi provided
helpful feedback and critical support throughout this project. We also
thank the Associate Editor and two anonymous for helpful comments. This
work was partially supported by NIH grant
\if0\blind{R01 AI085073 }\fi \if1\blind{BLINDED }\fi and the Lower Cape
Fear River Program. The content is solely the responsibility of the
authors and does not necessarily represent the official views of the
National Institutes of Health.} \\ 
      Department of Biostatistics, University of North Carolina Chapel Hill \\ 
      \\  \and 
        Michael G. Hudgens  \\ 
      Department of Biostatistics, University of North Carolina Chapel Hill \\ 
      \\  \and 
        Michael A. Mallin  \\ 
      Center for Marine Science, University of North Carolina Wilmington \\ 
      \\ }
  \maketitle
} \fi 

\if1\blind
{
  \bigskip
  \bigskip
  \bigskip
  \begin{center}
    {\LARGE\bf Downstream Effects of Upstream Causes}    
\end{center}
  \medskip
} \fi } \fi

\bigskip
\begin{abstract}
The United States Environmental Protection Agency considers nutrient
pollution in stream ecosystems one of the U.S.' most pressing
environmental challenges. But limited independent replicates, lack of
experimental randomization, and space- and time-varying confounding
handicap causal inference on effects of nutrient pollution. In this
paper the causal g-methods are extended to allow for exposures to vary
in time and space in order to assess the effects of nutrient pollution
on chlorophyll \emph{a} -- a proxy for algal production. Publicly
available data from North Carolina's Cape Fear River and a simulation
study are used to show how causal effects of upstream nutrient
concentrations on downstream chlorophyll \emph{a} levels may be
estimated from typical water quality monitoring data. Estimates obtained
from the parametric g-formula, a marginal structural model, and a
structural nested model indicate that chlorophyll \emph{a}
concentrations at Lock and Dam 1 were influenced by nitrate
concentrations measured 86 to 109 km upstream, an area where four major
industrial and municipal point sources discharge wastewater.
\end{abstract}

\noindent%
\vfill

\newpage
\spacingset{1.45} 

\section{Introduction}\label{introduction}

Nutrient pollution of U.S. streams costs billions of dollars each year
\citep{dodds2009}. The EPA calls reducing nutrient pollution in U.S.
waterways a ``high priority'' \citep{epa2015} and acknowledges that
Nitrogen-Phosphorous (NP) pollution is a causal factor in algal blooms.
However, the EPA's 2015 report also notes that since many factors may
contribute to a harmful algal bloom (HAB), ``it is often difficult or
impossible to say \emph{how much more} likely an HAB is because of
nutrient pollution.'' Lack of experimental manipulation and small sample
sizes are among many potential pitfalls in making causal inferences
using stream surveillance data \citep{norton2014}. The EPA's Causal
Analysis/Diagnosis Decision Information System (CADDIS) outlines a
reasoned, methodical process for assessing causality in stream
ecosystems \citep{norton2009}. \citet{suter2002}, among the primary
developers of CADDIS, state that data analysis methods in causal
assessments ``should be selected to best illuminate the association
given the amounts and types of data available.'' In this paper, a
potential outcome (or counterfactual) approach is considered for drawing
inference about the causal effects of nutrient pollution on stream
ecosystems.

Data on North Carolina's Cape Fear River is analyzed as a case study.
This is a large Piedmont-Coastal Plain system that is representative of
many riverine systems from Virginia south through North Florida
\citep{dame2000}. Formerly considered a moderately productive river
\citep{kennedy2008}, in 2009 it began experiencing harmful algal blooms
consisting of the cyanobacterium (blue-green alga) \emph{Microcystis
aeruginosa} near Lock and Dam 1 (LD1) that reappeared periodically
through 2012 \citep{isaacs2014}. Freshwater algal blooms are often
stimulated by phosphorus (P) loading \citep{howarth2006}, but in Coastal
Plain rivers and streams, algal blooms are largely stimulated by
nitrogen (N) loading \citep{mallin2004, dubbs2008}. In North Carolina
\citep{ncdenr2005} as well as many states and provinces, regulatory
agencies regularly monitor concentration of the algal pigment
chlorophyll \emph{a} as a proxy for algal bloom strength. Long-term
monitoring of this river by state-certified coalitions, including the
Lower Cape Fear River Program and Middle Cape Fear Coalition, has
provided a data set of nutrients, chlorophyll \emph{a}, and other water
quality parameters for the middle and lower river, where the blooms are
concentrated.

This paper shows that causal effects of upstream nutrient concentrations
on downstream chlorophyll \emph{a} can be estimated from observational
water quality data. Correlation analyses or regression techniques, while
invaluable for exploring associations within an ecosystem, do not
typically estimate causal effects. With publicly available watershed
monitoring data, we assess causal effects of nutrient concentrations
measured upstream of LD1 on chlorophyll \emph{a} levels at LD1 by
adapting the causal g-methods
\citep{robins2009estimation, hernanrobins2016}. Originally developed for
assessing the effect of a time-varying exposure, here the g-methods are
extended to the setting where exposure varies in both time and space. In
particular, the causal models allow for spatial interference
\citep{verbitsky2012, di2016policy} in the sense that exposure (nutrient
concentration) at one location may affect the outcome (chlorophyll
\emph{a}) at another location. Inference about parameters of marginal
structural models, the parametric g-formula, and structural nested
models which accommodate the spacetime interference structure of a
stream ecosystem is considered using estimating equation theory
\citep{stefanski2002}, with small sample adjustments \citep{fay2001} to
account for limited independent replicates.

The paper is organized as follows. Section
\ref{motivation-materials-and-notation} motivates the analysis and
describes the available data on the Cape Fear River. Section
\ref{causal-inference-from-upstream-to-downstream} introduces potential
outcomes, key assumptions, and the target estimand. A graphical
representation of the model assumptions using a Single World
Intervention Template \citep{richardson2013single} is also presented.
The g-methods are presented in Section
\ref{estimation-of-causal-effects} along with small sample variance
corrections. The simulation study in Section \ref{simulation-study}
validates and compares statistical properties of the g-methods. The Cape
Fear River data are analyzed in Section \ref{cape-fear-river-analysis}.
Finally, we discuss our findings and their limitations in Section
\ref{discussion}. The Supplementary Material contains the code and data
necessary to replicate the analyses plus additional mathematical and
analysis details.

\section{Motivation, materials, and
notation}\label{motivation-materials-and-notation}

\subsection{Cape Fear River nutrient pollution and algal
blooms}\label{cape-fear-river-nutrient-pollution-and-algal-blooms}

During the summers of 2009-2012, algal blooms unprecedented in scale and
composition occurred near LD1 near Kelly, NC. \citet{isaacs2014}
reported that samples collected from these blooms in 2009 and 2012
consisted predominantly of toxic \emph{Microcystis aeruginosa}
cyanobacteria. The multi-stakeholder watershed action plan for the Cape
Fear River identifies blue-green algae, \emph{M. aeruginosa} in
particular, as a significant threat to the river ecosystem
\citep{capefear2013}. Over 2 million people rely on drinking water from
the Cape Fear watershed, and algal blooms have impacted taste and odor
from some water treatment plants \citep{ahuja2013}. Brunswick County, in
southeastern North Carolina, obtains some of its drinking water directly
from the river near LD1. Taste and odor problems arising from the
cyanobacterial blooms forced the water utility to increase its level of
water treatment, at significant cost, to produce acceptable drinking
water. Thus, causes of the recent degradation in Cape Fear River water
quality are key management concerns.

\begin{figure}
\centering
\caption{The map shows the extent of the Cape Fear watershed within the political boundaries of North Carolina, as well as the region of interest for this study. The algal blooms generally occurred near LD1. This study examines causal relationships between nutrient concentration measured at upstream sampling locations (open triangles) on chlorophyll \textit{a} at LD1.}
\label{fig:capefear}
\includegraphics{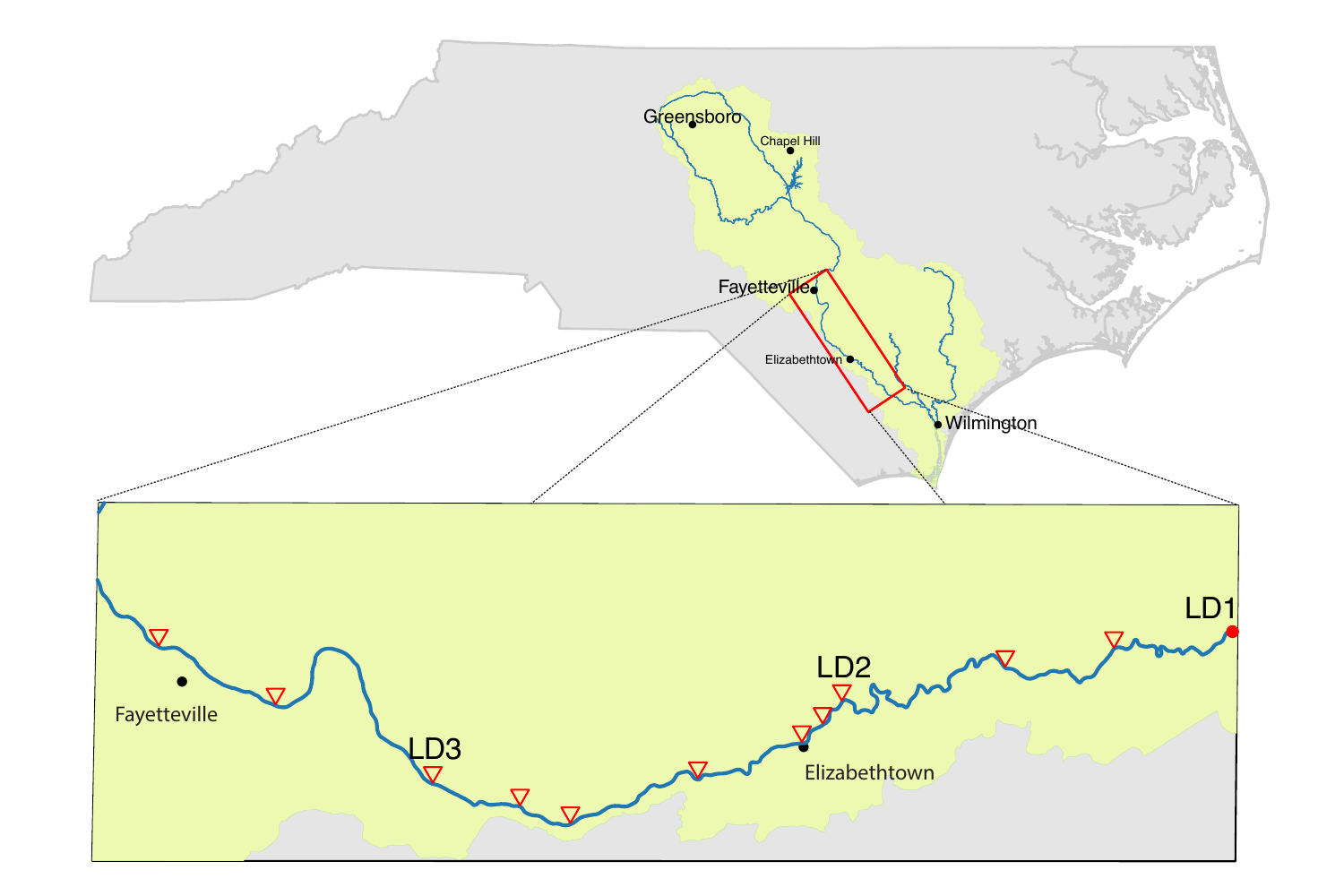}
\end{figure}

The 9000 square mile Cape Fear watershed is contained entirely within
the political boundaries of North Carolina, extends from Greensboro to
Wilmington, and includes parts of Durham and Chapel Hill. The Cape Fear
River forms at the confluence of the Haw and Deep Rivers and once
supported rich fisheries of anadromous fish \citep{capefear2013}. Figure
\ref{fig:capefear} shows the extent of the Cape Fear watershed and the
area of interest for this study, the section of river from Fayetteville
to LD1.

The Nature Conservancy of North Carolina obtained coalition-produced,
state-certified data consisting of monthly measurements from locations
throughout the Cape Fear basin from July 1996 through June 2013. Prior
to 1999, chlorophyll \emph{a} was not consistently measured at LD1.
Since large blooms at LD1 were reported mainly during summer months, we
focused our analysis on observations from June, July, August, and
September of 1999 to 2012 from the main stem of the Cape Fear River
upstream of LD1. The data include concentration measurements of four NP
compounds (all in mg/L): nitrate (NO\(_3\)), ammonia (NH\(_3\)), total
Kjeldahl nitrogen (TKN), and phosphorous (P).

\subsection{Associations of nutrients and LD1
chlorophyll}\label{associations-of-nutrients-and-ld1-chlorophyll}

\begin{figure}
\centering
\caption{Spearman's correlation coefficients between nutrient levels at upstream sampling locations and LD1 chlorophyll \textit{a}, using observations from June-September of 1999-2012. Open circles indicate the maximum correlation for a nutrient within this reach of the river.}
\label{fig:cfrcorrelations}
\includegraphics{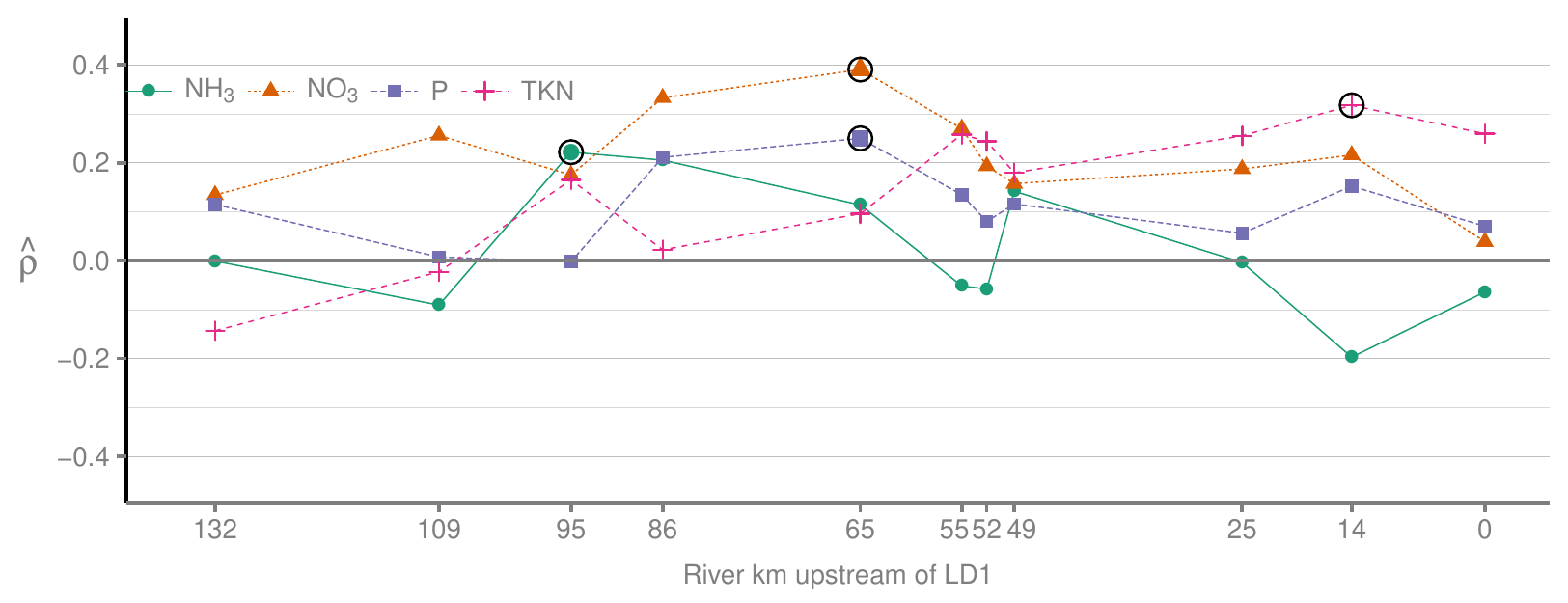}
\end{figure}

A simple correlation analysis shows generally positive associations
between upstream nutrients and chlorophyll \emph{a} levels at LD1.
Figure \ref{fig:cfrcorrelations} plots Spearman's correlation
coefficients between nutrient concentrations at sampling locations
within the study region and LD1 chlorophyll \emph{a}. Each nutrient has
a slightly different trajectory over the course of the river, but with
the exception of TKN, the correlation peaks between 65 and 95 river
kilometers upstream of LD1. These associations suggest a relationship
between upstream nutrient levels and LD1 chlorophyll. Our goal in this
paper is to adjust for confounding to determine to what degree the
upstream nutrients cause changes in LD1 chlorophyll \emph{a}.

\subsection{A mathematical
description}\label{a-mathematical-description}

Let \(i = 1, \dots, m\) index independent replicates; for the Cape Fear
data, \(m = 14\) corresponding to the years 1999 to 2012. We assume that
clusters of summer months are sufficiently far enough apart in time to
be considered independent. That is, observations from June to September
of year \(i\) are independent of the same set of observations in year
\(i' \neq i\), but observations within a year may be correlated. This is
the ``time-slices'' approach recommended by CADDIS for relieving
temporal autocorrelation\footnote{\url{https://www.epa.gov/caddis-vol4/caddis-volume-4-data-analysis-basic-principles-issues}}.
For a generic random variable \(W\), the indexing \(W_{ist}\) is used
where \(i\) indicates year, \(s = 1, \dots, n_s\) indicates the sampling
locations, ordered from upstream to downstream, and
\(t = 1, \dots, n_t\) indicates the month (e.g., \(t = 1\) denotes
June). In the following, the \(i\) notation is dropped where convenient.
In general capital letters denote random variables and lower case
letters denote indices, constants, or possible values of random
variables. The notation \(\operatorname{f}_w\) or
\(\operatorname{f}(w)\) is used as the probability density or mass
function for a random variable \(W\).

Observed values of chlorophyll \emph{a} (\(\mu\)g/L), the outcome of
interest, are \(\mbox{log}_2\) transformed and denoted as \(Y_{st}\).
For simplicity, chlorophyll \emph{a} is referred to as chlorophyll in
the following. The effect of each nutrient is considered separately in
our analysis, and nutrient exposure is generically denoted as
\(A_{st}\). Other covariates measured concurrently with nutrient
concentrations include the date and time of measurements, water
temperature (\(^{\circ}\)C), dissolved oxygen, pH, and turbidity. In
addition to covariates recorded in the water quality data set, daily
mean discharge data from stream gauges located at the William O'Huske
Lock and Dam (LD3) and LD1 were downloaded from USGS and converted to
m\(^3\)/s. Discharge values were linearly interpolated based on river
distance for sampling locations between the gauges. For each location,
the average of the mean daily discharge from the same date as the water
quality measurements plus the two prior days was used in the analysis.
Let \(L_{st}\) denote covariates measured at location \(s\) in month
\(t\). Let \(O_{st} = \{ Y_{st}, A_{st}, L_{st} \}\) denote the observed
random variables at location \(s\) in month \(t\). For any variable
\(W_{st}\), let the \(s \times t\) matrix \(\overline{W}_{st}\) denote
the variable's history for all locations upstream to and including
location \(s\), plus all time points prior to and including time point
\(t\).

\section{Causal inference from upstream to
downstream}\label{causal-inference-from-upstream-to-downstream}

Let \(Y_{s^{\star}t}(\overline{a}_{st})\) be the potential value of
\(\log_2\) chlorophyll at location \(s^{\star}\) in month \(t\) had the
exposure history been \(\overline{a}_{st}\), for \(s < s^{\star}\). By
causal consistency \citep{pearl2010},
\(Y_{s^{\star} t}(\overline{a}_{st}) = Y_{s^{\star} t}\) when
\(\overline{A}_{st} = \overline{a}_{st}\). Define the average potential
outcomes for a location of interest \(s^{\star}\) over months
\(t = 1, \dots, n_t\) as
\(\operatorname{E}\{ [Y_{s^{\star} 1}(\overline{a}_{s1}), Y_{s^{\star}2}(\overline{a}_{s2}), \dots, Y_{s^{\star}{n_t}}(\overline{a}_{s{n_t}})] ^{\intercal} \} = \operatorname{E}[Y_{s^{\star}}(\overline{a}_s)]\).
In the analysis, \(s^{\star} = 3\) corresponds with LD1.

\subsection{Effects of interest}\label{effects-of-interest}

Nutrient effects of interest to policymakers or scientists (the
estimands) can be stated in terms of functions of average potential
outcomes. For example, what difference would be expected, on average, in
LD1 chlorophyll levels if NH\(_3\) exposure at the upstream points LD3
and LD2 was set to be above, rather than below, a certain threshold
during the month of June (\(t = 1\))? Letting \(s_1\) correspond to LD3
and \(s_2\) correspond to LD2, this estimand is, in the notation defined
above,
\(\operatorname{E}[Y_{3 1}((a_{1 1}, a_{2 1})^{\intercal}) - Y_{3 1}((a'_{1 1}, a'_{2 1})^{\intercal})]\),
where \(a_{11}\) and \(a_{21}\) indicate NH\(_3\) exposure at LD3 and
LD2 above the NH\(_3\) threshold and \(a_{11}'\) and \(a_{21}'\)
indicate NH\(_3\) levels below the threshold.

Consider the estimand which measures the effect of setting nutrient
concentrations at two upstream locations, \(s_1\) and \(s_2\), on LD1
chlorophyll averaged across \(n_t\) months, i.e.,

\begin{equation}
\label{estimand1}
\mu(a_t, a'_t) = \frac{1}{n_t} \sum_{t = 1}^{n_t} \operatorname{E}\left\{ Y_{3t}(0_{t - 1}:a_t) - Y_{3t}(0_{t - 1}:a'_t) \right\}, 
\end{equation}

\noindent where \(0_t\) is a \(2 \times t\) matrix of zeros defined as
the empty set when \(t = 0\), \(a_{t} = (a_{1t}, a_{2t})^{\intercal}\),
and \(R:Q\) indicates the concatenation of matrices \(R\) and \(Q\).
Note \(0\) without a subscript denotes the scalar zero. The estimand
\eqref{estimand1} is defined in general for any two exposure settings
\(a_t\) and \(a'_t\). In the Cape Fear River analysis, exposure is
defined as a 2-tuple of binary variables both being above
(\(a_t = (1, 1)^{\intercal}\)) or both below
(\(a'_t = (0, 0)^{\intercal}\)) cutpoints specified in Section
\ref{cape-fear-river-analysis}. For brevity, \(\mu(a_t, a'_t)\) is
denoted \(\mu\).

The estimand \eqref{estimand1} characterizes the average effect on LD1
chlorophyll when intervening at two upstream locations simultaneously.
This parameter is of interest to the community of scientists working on
the Cape Fear River who want to understand the effects of nutrient
concentrations from different upstream locations on LD1 chlorophyll
during the summer when the toxic algal blooms generally occurred.
Assessing the causal effect of exposures at two upstream locations
simultaneously requires adjusting for covariates that affect the
exposure and vary between the upstream locations. Not correctly
accounting for such covariates may result in biased inferences about the
nutrient effects.

\subsection{Single world intervention
graph}\label{single-world-intervention-graph}

\citet{richardson2013single} introduced single world intervention graphs
(SWIGs) to unify the graphical approach to causal inference \citep[e.g.,
see][]{pearl2009} and the more algebraic potential outcomes framework
\citep[e.g., see][]{rubin2005}. An important difference between the
approaches is the representation of potential outcomes. Algebraic
notation can easily distinguish between potential and observed outcomes
(\(Y(a)\) versus \(Y\)). Directed acyclic graphs (DAGs) do not
explicitly encode potential outcomes. SWIGs do.

\begin{figure}
\caption{This single world intervention template conceptualizes the upstream to downstream process. $A_{st}$ is an exposure of interest, and $a_{st}$ is a fixed value of the exposure. At location 2, space-varying covariates labeled $L^{\bullet}_{2t}$ are associated with the potential outcome $Y_{3t}(\overline{a})$, affect $A_{2t}$, and are affected by $A_{1t}$. $L_{2t}$ covariates without the space-varying confounding properties are labeled $L^{\circ}_{2t}$. Within nodes with random variables that depend on past history, $\overline{a}$ is a generic history whose contents depend on $s$ and $t$ as described in the main text. This SWIT does not include all possible arrows. For example, $a_{1t} \to L_{2t}(\overline{a})$ would imply an effect of the exposure from the previous month and location. Since the strongest effects should occur within a month, temporal arrows are limited to the effects of covariates and exposures at the same node in the following month.}
\label{fig:swit}
\includegraphics{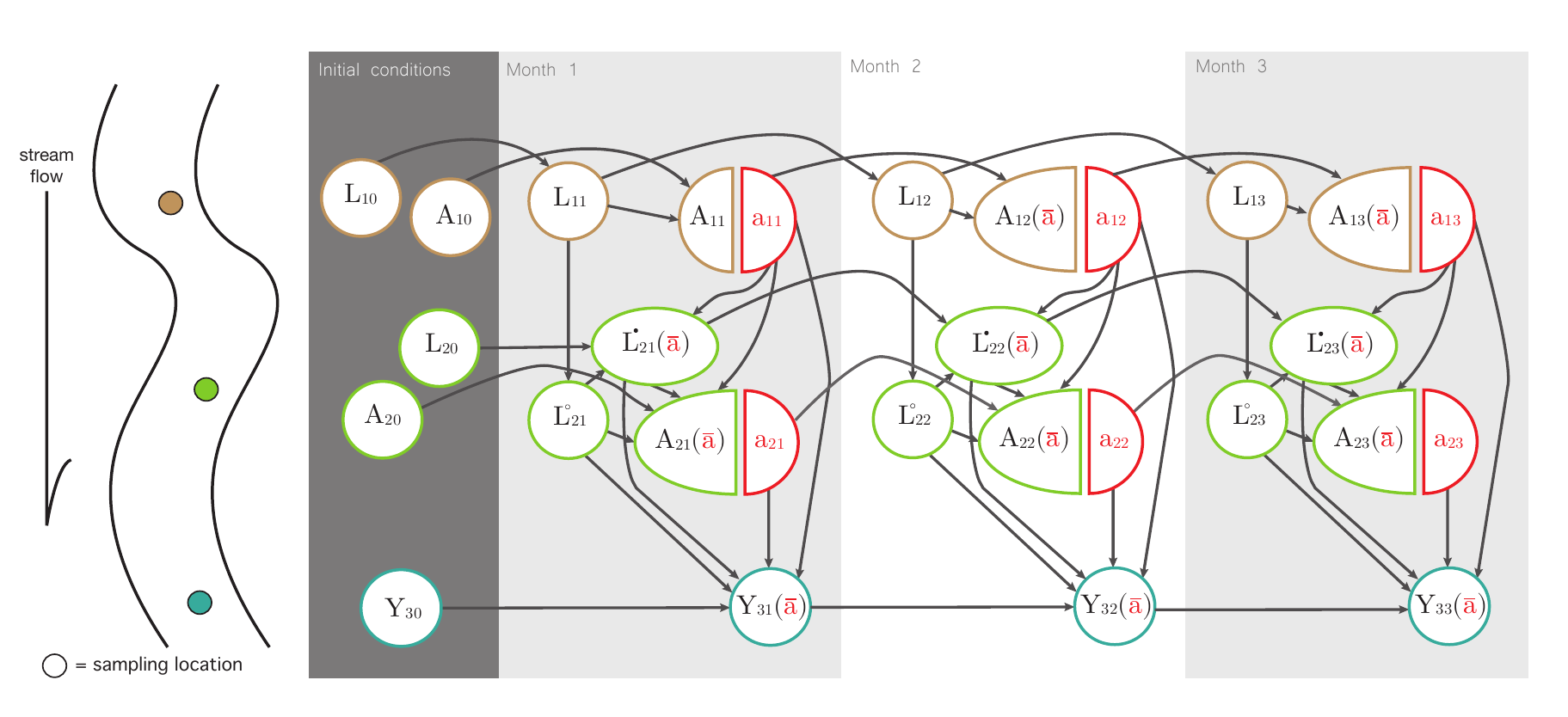}
\end{figure}

Reading a SWIG is similar to reading a DAG. Nodes represent variables
and edges suggest causal relationships between nodes (Figure
\ref{fig:swit}). In a SWIG, however, intervention nodes are transformed
by a node-splitting operation. Instead of a single \(A_{11}\) node as in
a DAG, the \(A_{11}\) semicircle represents the random variable for
exposure at location \(1\) at time \(1\). The \(a_{11}\) semicircle
represents the fixed setting of the exposure (possibly contrary to fact)
at the same spacetime point. Figure \ref{fig:swit} is technically a
single world intervention template (SWIT), not a SWIG. SWITs are a
graphical template for a set of exposure levels, whereas SWIGs represent
the graph for a single exposure level. We assume the SWIGs have the same
form for all exposures and all their levels, hence a single SWIT
describes the SWIGs for all exposure levels.

\section{Estimation of causal
effects}\label{estimation-of-causal-effects}

Time-varying confounding may introduce bias if not accounted for in
estimation. \citet{robins2009estimation} describe three g-methods for
estimating causal effects from observational data in the presence of
time-varying confounding: the parametric g-formula, fitting marginal
structural models (MSMs) using inverse probability weighting, and
g-estimation of structural nested models (SNMs). This section describes
extensions of these g-methods to the spacetime setting.

\subsection{Causal assumptions}\label{causal-assumptions}

The causal effect \(\mu\) can be identified by the distribution of the
observable random variables by considering the structure of a stream
(represented by Figure \ref{fig:swit}) as a sequentially and
conditionally randomized experiment. Given (i) covariate values up to
and including location \(s\) and month \(t\) and (ii) values of the past
exposure(s) prior to location \(s\) and month \(t\), \(A_{st}\) is
assumed to be independent of the potential outcomes. That is, the
covariate and exposure histories must block all back-door paths between
\(A_{st}\) and \(Y_{s^{\star} t}(\overline{a}_{st})\) \citep{pearl2009},
which implies conditional independence, commonly referred to as the
strong ignorability or no unmeasured confounding assumption:

\begin{equation}
\tag{A1}
\label{nuc}
Y_{s^{\star} t}(\overline{a}_{st}) \perp A_{st} | \overline{L}_{st}, \overline{A}^{\dagger}_{st},  
\end{equation}

\noindent where
\(\overline{A}^{\dagger}_{st} = \overline{A}_{st} \setminus \{ A_{st} \}\).

For the g-formula and MSMs, identification of causal effects also
depends on a positivity assumption
\(\operatorname{f}(a_{st} | \overline{O}^{\dagger}_{st} = \overline{o}^{\dagger}_{st}) > 0\)
for all \(\overline{o}^{\dagger}_{st}\) such that
\(\operatorname{f}(\overline{o}^{\dagger}_{st}) > 0\) where
\(\overline{O}_{st}^{\dagger} = \overline{O}_{st} \setminus \{A_{st} \}\).
That is, each level of exposure must have some non-zero probability of
occurring at all spacetime points for all possible covariate and
exposure histories.

These assumptions are needed to identify causal effects
nonparametrically. In many applications, as in ours, common
finite-dimensional parametric models such as linear or logistic
regression are employed to model aspects of the distribution of
observable random variables. These models must be correctly specified in
order for the resulting inferences to be valid.

\subsection{Parametric g-formula}\label{parametric-g-formula}

The g-formula is a mathematical identity which relates the distribution
of counterfactuals to the distribution of the observable random
variables \citep{robins1986a-new-approach, robins2009estimation}. For
example, using the g-formula, the counterfactual mean can be expressed
as:

\begin{equation}
\label{gfm}
\operatorname{E}[Y_{s^{\star} t}(\overline{a}_{st})] = \int_{\overline{l}_{st}}  \operatorname{E}[Y_{s^{\star}t} | \overline{A}_{st} = \overline{a}_{st}, \overline{L}_{st} = \overline{l}_{st}] \operatorname{f}_{\overline{l}_{st}} \mathrm{d} \overline{l}_{st} .
\end{equation}

\noindent where
\(\operatorname{f}_{\overline{l}_{st}} = \prod_{j = 1}^{s} \prod_{k = 1}^{t} \operatorname{f}_{l_{jk}| \overline{l}_{j-1, k-1}, \overline{a}_{j-1, k-1}}\).
In practice, the mean model
\(\operatorname{E}[Y_{s^{\star}t} | \overline{A}_{st} = \overline{a}_{st}, \overline{L}_{st} = \overline{l}_{st}]\)
and conditional densities or mass functions
\(\operatorname{f}_{l_{jk}| \overline{l}_{j-1, k-1}, \overline{a}_{j-1, k-1}}\)
are not known, and estimated values
\(\hat{\operatorname{E}}[Y_{s^{\star}t} | \overline{A}_{st} = \overline{a}_{st}, \overline{L}_{st} = \overline{l}_{st}]\)
and \(\hat{\operatorname{f}}_{\overline{l}_{st}}\) are plugged into
\eqref{gfm} to estimate
\(\operatorname{E}[Y_{s^{\star} t}(\overline{a}_{st})]\). Though these
quantities may be estimated nonparametrically for a single spacetime
point, a parametric approach may be necessary to estimate more
complicated quantities such as \(\mu\). In both the analysis and
simulations presented below, the mean model was parameterized as a
linear model with main effects only for \(A_{2t}\), \(A_{1t}\), and
\(L_{2t}\), with corresponding parameters \(\beta_1^g\), \(\beta_2^g\),
and \(\beta^g_4\), respectively. Let \(\gamma_3^g\) be the parameter
corresponding to \(A_{1t}\) in a simple linear model for the mean of
\(L_{2t}\). When the exposure settings are binary where \(a_{st} = 1\)
and \(a'_{st} = 0\) for all \(t\), then
\(\mu = {n_t}^{-1} \sum_{t = 1}^{n_t} \left\{ \beta^{g}_1 + (\beta^{g}_2 + \beta^{g}_4\gamma^g_3) \right\} = \beta^{g}_1 + \beta^{g}_2 + \beta^{g}_4 \gamma^g_3.\)

\noindent The Supplementary Material contains the algebraic details.
Maximum likelihood is used to estimate the model coefficients. The
coefficient estimates are then plugged into \eqref{gfm} to obtain the
estimator of \(\operatorname{E}[Y_{s^{\star} t}(\overline{a}_{st})]\).

One drawback to the parametric g-formula is the g-null paradox, wherein
if the null hypothesis of no treatment (exposure) effect is true,
plugging standard parametric models into \eqref{gfm} will result in
rejection of this null \citep{robins2009estimation} as sample size
increases. The inferential approaches in the next two sections do not
suffer the g-null paradox.

\subsection{Marginal structural model}\label{marginal-structural-model}

Marginal structural models posit a parametric relationship between an
exposure history and a counterfactual outcome. Consider the following
MSM:

\begin{equation}
\label{msm}
\operatorname{E}[ Y_{s^{\star}t}(\overline{a}_{st}) ] = \beta^{m}_{0t} + \beta^{m}_1 a_{st} + \beta^{m}_2 a_{s - 1,t}  .
\end{equation}

\noindent Each month may have a distinct intercept \(\beta^{m}_{0t}\),
but the counterfactual mean depends only on exposure at two upstream
locations during the same month. From \eqref{msm},
\(\mu = \beta^{m}_1 + \beta^{m}_2\). Parameters in MSMs can be estimated
consistently using inverse probability weighting methods
\citep{hernan2000marginal}. We use the stabilized inverse probability
weight where each observed outcome is weighted by:

\begin{equation}
\label{sw}
SW_{st} = \prod_{j = 1}^s \prod_{k = 1}^t \frac{\operatorname{f}(a_{jk}| \overline{A}_{jk}^{\dagger} = \overline{a}_{jk}^{\dagger})} {\operatorname{f}(a_{jk} | \overline{O}_{jk}^{\dagger} = \overline{o}_{jk}^{\dagger} )} .
\end{equation}

\noindent The product is taken across the dimensions of space \(s\) and
time \(t\) as opposed to a single dimension as in
\citet{robins2000marginal}. Logistic regression is used to estimate
\(\operatorname{f}(a_{st}| \overline{A}_{st}^{\dagger} = \overline{a}_{st}^{\dagger})\)
and
\(\operatorname{f}(a_{st} | \overline{O}^{\dagger}_{st} = \overline{o}^{\dagger}_{st})\)
(see Supplementary Materials for details). Weighting observed outcomes
by \eqref{sw}, generalized estimating equations (GEE) \citep{liang1986}
with an independence working correlation matrix are used to estimate
\(\beta^{m} = (\beta^m_0, \beta^m_1, \beta_2^m)\).

\subsection{Structural nested (mean)
model}\label{structural-nested-mean-model}

Instead of modeling counterfactual means from which causal contrasts are
then derived, structural nested models directly model a causal effect
\citep{robins1994correcting}. In general, SNMs model the effect of
removing treatment (exposure) within strata \(l\),
\(\operatorname{E}[Y(a) - Y(0) | L = l]\). \citet{vansteelandt2014}
describe several advantages of SNMs over MSMs. For one, target
parameters in SNMs are identified without the positivity assumption. The
asymptotic variance of IPW estimators also tends to be more sensitive to
misspecification of the model(s) compared to G-estimators of SNM
parameters \citep{vansteelandt2014}. SNMs can also be used to test the
null hypothesis of no effect for \emph{any} treatment regime, which MSMs
cannot do \citep{robins2000msmvssnm}.

The Cape Fear River analysis uses the following structural nested mean
model:

\begin{equation}
\label{snm}
\left[ \begin{matrix}
\vdots \\
\operatorname{E} \left\{ Y_{3t}\begin{pmatrix}
\begin{array}{cc} 
\multirow{2}{*}{$0_t$} & a_{1t} \\
  & a_{2t} 
\end{array}
 \end{pmatrix} - Y_{3t}
\begin{pmatrix}
\begin{array}{cc} 
\multirow{2}{*}{$0_t$} & a_{1t} \\ 
 & 0  
\end{array}
\end{pmatrix} \bigg| \overline{L}_{2t} = \overline{l}_{2t} \right\} \\
\operatorname{E} \left\{ Y_{3t}\begin{pmatrix}
\begin{array}{cc}
\multirow{2}{*}{$0_t$}  & a_{1t} \\  & 0  
\end{array}
\end{pmatrix} - Y_{3t}\begin{pmatrix} 
\begin{array}{cc}
\multirow{2}{*}{$0_t$} & 0 \\ 
& 0 
\end{array}
\end{pmatrix} \bigg| \overline{L}_{1t} = \overline{l}_{1t} \right\} \\
\vdots
\end{matrix}
\right] = \begin{bmatrix} 
\vdots \\
\operatorname{h}_{1t}(\overline{a}_{t}, \overline{l}_{2t}; \beta^{s}) \\
\operatorname{h}_{2t}(\overline{a}_{t}, \overline{l}_{1t}; \beta^{s}) \\
\vdots \\
\end{bmatrix} = \begin{bmatrix} 
\vdots \\
\beta^{s}_1 a_{2t} \\
\beta^{s}_2 a_{1t} \\
\vdots \\
\end{bmatrix}. 
\end{equation}

\noindent For \(n_t = 4\), \eqref{snm} has dimension \(8 \times 1\), as
each month \(t\) has two \(\operatorname{h}\) functions. The
\(\operatorname{h}\) function corresponds to a ``blip down'' process
\citep{vansteelandt2014}, removing the effect of treatment one spatial
location at a time. The first \(\operatorname{h}_{1t}\) ``blips out''
and quantifies the effect of \(a_{2t}\), and \(\operatorname{h}_{2t}\)
quantifies the effect of \(a_{1t}\). This SNM assumes that
\(\operatorname{h}\) does not depend on \(l\); that is, the causal
contrast does not include interactions between exposure and covariates.
Thus, \(\beta^{s}_1 + \beta^{s}_2 = \mu\). While in general a SNM can
include interactions between exposures and spacetime-varying confounding
covariates, we specify \eqref{snm} with no interactions in order for
\(\mu\) to equal a simple linear function of the SNM parameters.

To estimate \(\beta^{s}\), a vector \(U_t(\beta^{s})\) is constructed
whose mean value, given \(\overline{L}_{st}\) and \(\overline{A}_{st}\),
equals the mean outcome that would have been observed had treatment been
removed:

\[ 
U_t(\beta^{s}) = \begin{pmatrix} 
U_{1t}(\beta^{s}) \\
U_{2t}(\beta^{s})
\end{pmatrix} =
\begin{pmatrix} 
Y_{3t} - \beta^{s}_1 A_{2t} \\
Y_{3t} - (\beta^{s}_1 A_{2t} + \beta^{s}_2 A_{2, t - 1})
\end{pmatrix} .
\]

\noindent Using a modified version of equation (33) in
\citet{vansteelandt2014}, the solution to the following estimating
equations is a consistent estimator for \(\beta^{s}\):

\begin{equation}
\label{snmee}
\sum_i \sum_{s = 1}^{n_s} \sum_{t = 1}^{n_t}  \left[ \operatorname{d}_s(\overline{L}_{ist}, \overline{A}_{ist}) - \operatorname{E}\{\operatorname{d}_s(\overline{L}_{ist}, \overline{A}_{ist}) | \overline{L}_{ist}, \overline{A}_{ist}\} \right] \left[ U_{ist}(\beta^{s}) - \operatorname{E}\{U_{ist}(\beta^{s}) | \overline{L}_{ist}, \overline{A}_{ist} \} \right] = 0,
\end{equation}

\noindent where \(\operatorname{d}_s\) is some arbitrary distance
function. Per the suggestion of \citet{vansteelandt2014}, we let
\(\operatorname{d}_s = \operatorname{E}\left\{ \partial U_{st}(\beta)/\partial \beta | \overline{L}_{st}, \overline{A}_{st}\right\}\).
The solution to \eqref{snmee} is called a G-estimator and has the
advantage of double robustness. That is, the estimator is consistent
when either the model for the transformed outcome
\(\operatorname{E}[U_{st}(\beta) | \overline{L}_{st}, \overline{A}_{st}]\)
or the exposure model (which is a component of
\(\operatorname{E}[\operatorname{d}_s(\overline{L}_{st}, \overline{A}_{st}) | \overline{L}_{st}, \overline{A}_{st}]\))
is correctly specified. Parametric regression models were used to model
both the outcome and exposure (linear and logistic regression,
respectively). In some cases, solving \eqref{snmee} yields a closed form
solution for \(\hat{\beta}^{s}\) as shown in the Supplementary Material,
which simplifies computations and makes the solution's uniqueness easy
to check.

\subsection{Estimating equation
inference}\label{estimating-equation-inference}

In each of the previous three sections, the g-formula \eqref{gfm}, MSM
\eqref{msm}, and SNM \eqref{snm} were specified such that parameter
estimates may be obtained by solving a set of unbiased estimating
equations. Therefore, under certain regularity conditions, the
estimators will be consistent and asymptotically normal, and the
empirical sandwich variance estimator can be used to consistently
estimate the asymptotic covariance matrix of the model parameter
estimators. In the case of the g-formula, the target estimand \(\mu\) is
a function of \(\beta^{g}_1\), \(\beta^{g}_2\), \(\beta^{g}_4\), and
\(\gamma^g_3\), so the estimated variance of \(\hat{\mu}\) can be
obtained using the delta method. For MSMs, observed outcomes for each
time point are weighted by estimated values of \eqref{sw}, and weighted
generalized estimating equations are used to obtain \(\hat{\beta}^{m}\).
Variance estimates for MSMs can be obtained by stacking the score
equations for the parametric models used to estimate the weights plus
the estimating equations corresponding to \eqref{msm} weighted by
\eqref{sw}. Point estimates in the SNM were obtained from the closed
form of \(\hat{\beta}^{s}\), while variance estimates were obtained by
stacking the score equations of both the outcome and exposure models
along with estimating equation \eqref{snmee}.

For all three methods, consistent variance estimators follow from
estimating equations (i.e., M-estimation) theory \citep{stefanski2002}.
Let \(\hat{\theta}\) be the estimator that solves the set of \(p\)
equations \(\sum_{i = 1}^m \operatorname{g}(O_{i}, \hat{\theta}) = 0\),
where \(\operatorname{g}\) is a vector of functions of length \(p\)
corresponding to the number of parameters in \(\theta\). From our causal
models, \(\theta\) contains the target parameters \(\beta\) plus any
nuisance parameters present in estimating the IP weights, outcome model,
or exposure model. The asymptotic covariance for \(\hat{\theta}\) is
\(\Sigma = A^{-1} B \{A^{-1}\}^{\intercal}/m\), where
\(A_i = \partial \operatorname{g}(O_i, \theta)/ \partial \theta\),
\(A = \operatorname{E}[A_i]\),
\(B_i = \operatorname{g}(O_i, \theta) \operatorname{g}(O_i, \theta)^{\intercal}\),
and \(B = \operatorname{E}[B_i]\). The empirical sandwich variance
estimator replaces the expectations with their empirical counterparts
and \(\theta\) with \(\hat{\theta}\); e.g.,
\(\hat{A} = m^{-1} \sum_{i = 1}^m \partial \operatorname{g}(O_i, \theta)/ \partial \theta |_{\theta = \hat{\theta}}\).

The empirical sandwich variance estimator is asymptotically consistent
but tends to underestimate the true variance in small samples
\citep{fay2001, li2014}. In the next section, we examine the bias
corrected estimator of \citet{fay2001} in simulations. The bias
corrected variance estimator replaces \(\hat{B}_i\) with
\(\hat{B}_i^{bc}(b) = \hat{H}_i(b) \hat{B}_i \hat{H}_i(b)^{\intercal}\)
to form
\(\hat{\Sigma}^{bc} = \hat{A}^{-1} \hat{B}^{bc}(b) \{\hat{A}^{-1}\}^{\intercal}\),
where
\(\hat{H}_i(b) = \{1 - \min(b, \{\hat{A}_i \hat{A}\}_{jj}) \}^{-1/2}\)
and \(\{\hat{A}_i \hat{A}\}_{jj}\) denotes the \(jj\)th element of
\(\hat{A}_i \hat{A}\) and
\(\hat{B}^{bc}(b) = \sum_i \hat{B}_i^{bc}(b)\). The constant \(b\) is
less than 1 and chosen by the analyst intended to prevent extreme
corrections when \(\hat{A}_i \hat{A}\) is close to 1. \citet{fay2001}
``arbitrarily'' set \(b = 0.75\). To explore the stability of our
variance estimates we used \(b = 0.1, 0.3\), and \(0.75\). Variance
estimates were used to construct Wald confidence intervals based on
either a normal distribution or a \emph{t} distribution with \(m\)
degrees of freedom.

\section{Simulation study}\label{simulation-study}

Based on the SWIT in Figure \ref{fig:swit}, we used the simcausal R
package \citep{simcausal} to simulate data for
\(m = 10, 15, 20, 25, \text{ and } 30\) years. For each \(m\), 24,000
data sets were generated according to the parametrization provided in
the Supplementary Material. The parameter values were based on estimates
from simple linear or logistic regressions using the Cape Fear data with
NO\(_3\) \textgreater{} 1 mg/L as the exposure. For each simulated data
set, estimates and 90\% confidence intervals of \(\mu\) were computed
using the causal g-methods, plus a naive GEE approach that ignores
space- and time-varying confounding. Code for the simulations is
available in the Supplementary Material.

Figure \ref{fig:simresults} shows the absolute bias of the four
estimators and empirical coverage of the corresponding confidence
intervals for each set of simulations. The horizontal shading highlights
bias \(< 0.01\), and the vertical shading highlights coverage within 1\%
of the nominal 90\% coverage. Each facet in the figure shows a different
correction to the variance estimator and distribution used in
constructing the Wald confidence intervals. For all three causal
methods, the absolute bias shrinks and the coverage improves as sample
size (i.e., number of years) increases. The bias for the MSM and
g-formula (GFM) methods is smaller compared to the SNM for all \(m\)
analyzed in Figure \ref{fig:simresults}. The SNM still has an average
absolute bias of about 0.01 even when \(m = 30\). In a secondary
simulation of 1000 data sets where \(m = 150\) (not shown), the SNM bias
shrunk to the same order of magnitude as the MSM and g-formula methods.
The naive GEE estimator is always biased, empirically illustrating why
methods that account for time- and/or space-varying confounding should
be used when such effects are present. For small \(m\), the unadjusted
variance estimator performs poorly, with Wald confidence intervals
covering in the 75 - 85\% range for all the methods (panels I1 and I5 in
Figure \ref{fig:simresults}). Larger values of \(b\) tend to overcorrect
the variance estimate, but confidence interval coverage with \(b = 0.1\)
approximates the nominal level.

\begin{figure}
\thisfloatpagestyle{empty}
\centering
\caption{Simulation results show absolute bias $|\hat{\mu} - \mu|$ on the y-axis. Each line shows results for a method from $m = 10$ (triangle) to $m = 30$ (square). The proportion of simulations where the 90\% confidence interval included the true value is shown on the x-axis. Each box shows results for the different methods of forming confidence intervals, with the columns defining the distribution and the row defining the form of the variance estimator. The bias is unaffected by corrections to the confidence intervals, hence the y-coordinates do not change across the boxes.}
\label{fig:simresults}
\includegraphics{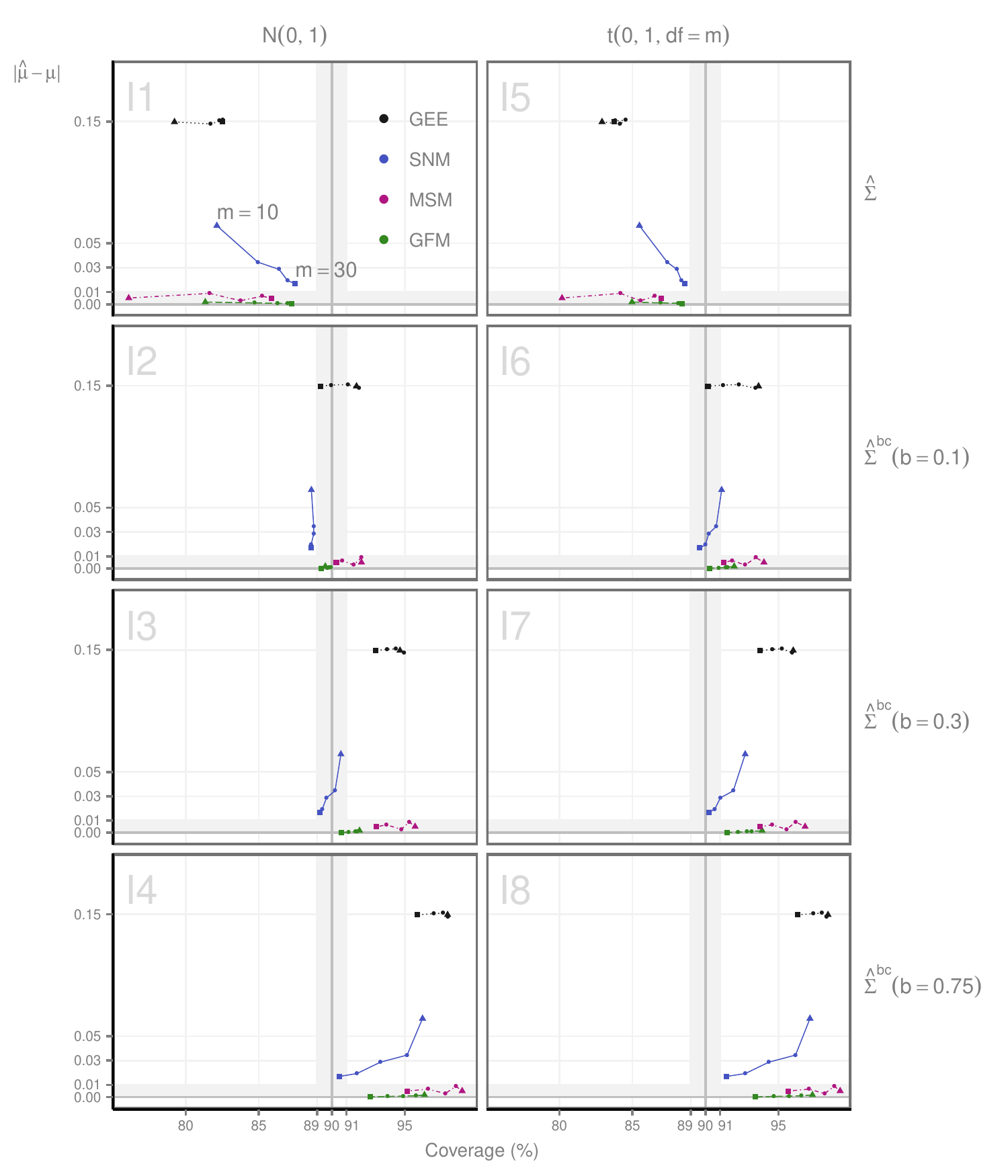}
\end{figure}

\section{Cape Fear River analysis}\label{cape-fear-river-analysis}

Beginning with the sampling location 132 km upstream near I-95 in
Fayetteville, we estimate \(\mu\) for a given NP species at that site
(\(s = 2\)) and the site just upstream (\(s = 1\)). That is, the causal
effect was estimated for setting an exposure at location A and location
B, then the effect of location B and location C, then the effect of
location C and D, and so on downstream until the last sampling location
upstream of LD1.

The methods described above treat exposures as binary, but the species
of NP are measured on a continuous scale. For each species and set of
upstream locations, the exposures were discretized using three cutpoints
based on the 1st, 2nd and 3rd quartiles at the two upstream locations
during April to October of 1999. The distribution for each NP species
varied over the course of the river, so a single river-wide cutpoint
would not be meaningful. At some locations, for example, all the
observed concentrations could be below a river-wide cutpoint. For each
nutrient, five space-varying confounding covariates were considered:
each of the other three nutrients, dissolved oxygen, or pH.

In summary, 40 causal effects were estimated (4 NP species \(\times\) 10
sets of \(s_2\) and \(s_1\) upstream locations) using each of the four
different analysis methods (g-formula, MSM, SNM, and GEE). On the
continuum fron exploratory to confirmatory science, we view these
results as exploratory causal analysis. For this reason, no control for
multiple comparisons was made, and in the spirit of a pilot study
\citep{lee2014the-statistical}, 90\% confidence intervals are presented.
Due partly to the small sample size and the simplified model (Figure
\ref{fig:swit}) of a complex ecological web, these results do not
confirm the magnitude and direction of causal effects; instead, these
results should both be interpreted in light of existing evidence and
inform future research.

The stability \citep[\emph{sensu}][]{rosenbaum2002observational} of
these results was assessed by estimating the effects for each of three
exposure cutpoints, five space-varying confounding covariates, and two
different parameterizations of the outcome and exposure models. The
stability analyses are summarized in the Supplementary Materials. Wald
confidence intervals and p-values were computed using the eight
combinations of distribution and variance estimator as in Figure
\ref{fig:simresults}. For each analysis method, the outcome and exposure
models were parameterized similarly to the simulations, with the
exception that \(L^{\circ}_{st}\) was not a single covariate and instead
both temperature and discharge were included in the models.

For the observations considered in our analysis, at most 5\% of the data
were missing for any given covariate. LD1 chlorophyll values were
missing for May 2000 and July 2004. These two missing values were
replaced by taking the average from the month prior and post. Of the 770
observations for the 11 upstream locations during May to September of
1999 to 2012, 5 NH\(_3\), 5 NO\(_3\), 22 TKN, 38 P, and 16 temperature
values were missing. No NH\(_3\) or NO\(_3\) were missing for June to
September. Where possible, missing covariate values at a location-month
were singly imputed in the following sequential manner. First, we
attempted to average the values immediately upstream and immediately
downstream during the same month. If either value immediately upstream
or immediately downstream was missing, then we averaged values from the
next site upstream and next site downstream during the same month. If
neither of these approaches imputed a value, then we averaged the prior
month and next month from the same location. This approach imputed
missing values for all nutrient covariates except P, which had missing
values for all summer months at all of our upstream locations in 2009.
We excluded 2009 in analyses involving P. The stability of this simple
approach to missing data was checked using a multiple imputation
procedure \citep{hmisc}, which is described in the Supplementary
Materials. The results in Figure \ref{fig:cfrresults} do not
substantively change with the multiple imputation.

Figure \ref{fig:cfrresults} shows causal effect estimates with
point-wise 90\% confidence intervals for the four nutrient species using
the median cutpoint for the exposure as described above. To be
conservative, the confidence intervals are based on
\(\hat{\Sigma}(b = 0.3)\). GEE and g-formula results were largely
similar, so we only show g-formula results. Although the GEE and
g-formula results were alike for these data, this will not be the case
in general as demonstrated in the simulation study in Section
\ref{simulation-study}. The space-varying covariate is P for the
nitrogen species, and it is NH\(_3\) for P. In this set of analyses, one
of the models for the terms in \eqref{sw} failed to converge when
estimating the MSM in four cases (three for NO\(_3\) and one TKN). These
are indicated by open points, which were interpolated from the other
values, unless the model failed for the location just upstream of LD1.
Confidence intervals are based on the bias corrected standard errors
(\(b = 0.3\)) and a \(t\)-distribution with \(m\) degrees of freedom.
Thick bars with stars above indicate confidence intervals which exclude
the null value of \(0\).

\begin{figure}
\thisfloatpagestyle{empty}
\centering
\caption{Results of causal analysis of Cape Fear River data from June-September of 1999-2012 for each of the four measured nutrients. Points are estimates of $\mu$. Vertical lines are 90\% confidence intervals with thicker lines highlighting intervals that do not cross zero. The intervals used in this plot use method I7 of Figure \ref{fig:simresults}. Open points are where the model fitting algorithm failed to converge.}
\label{fig:cfrresults}
\includegraphics{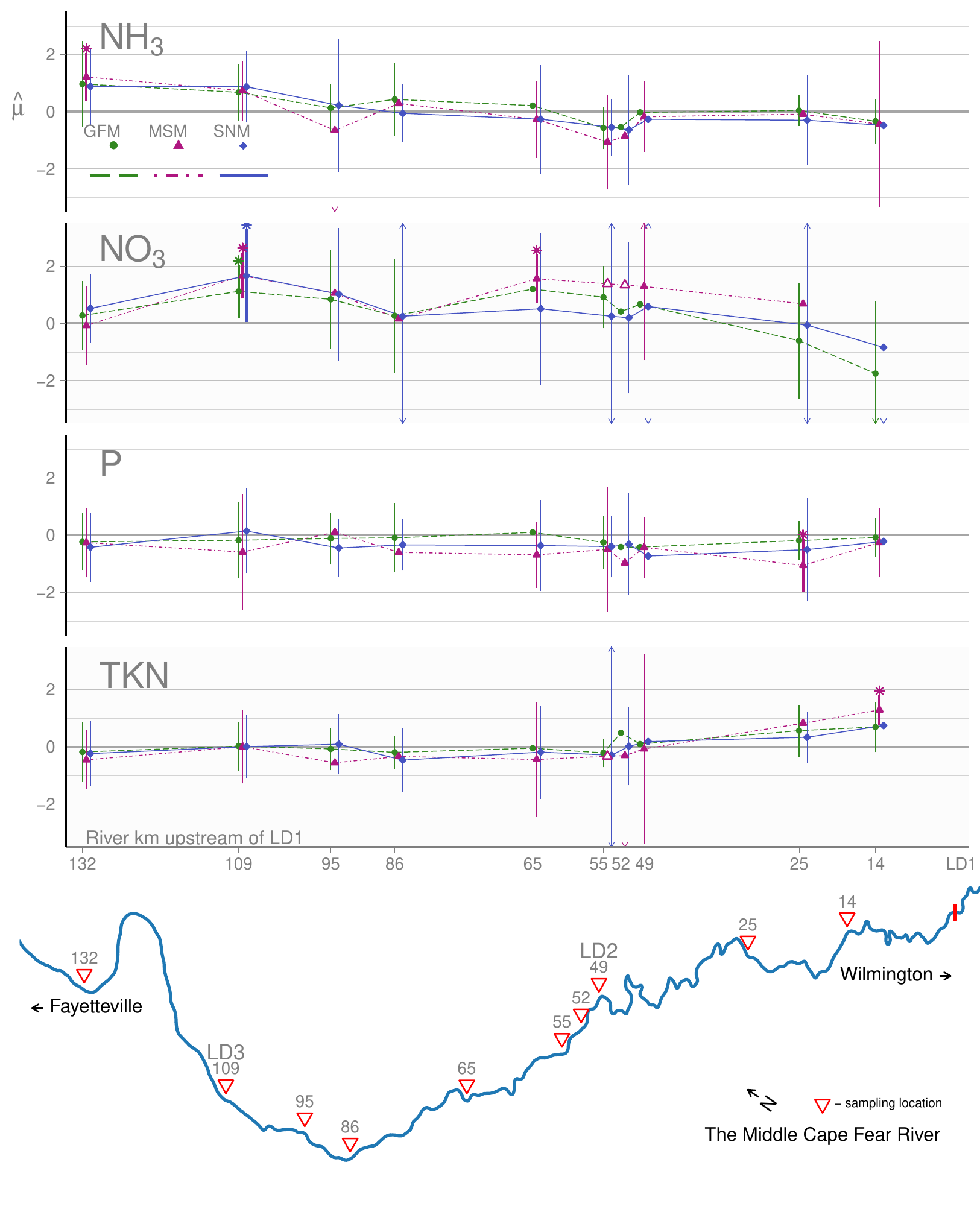}
\end{figure}

The point estimates of the three causal methods tend to have similar
results. Standard errors were also of similar magnitude, with the
exception of estimates when \(s_2\) is LD2 where the standard errors for
the SNM tended to be uninformative. All three methods indicate a
statistically significant effect of NO\(_3\) when \(s_2\) is the LD3
sampling location, 109 kilometers upstream of LD1. The point estimates
at this location for NO\(_3\) were 1.12 for the g-formula and 1.67 for
the MSM and SNM, implying a 2- to 3-fold increase in LD1 chlorophyll
when NO\(_3\) is above 0.38 mg/L at both the location 109km upstream
(LD3) and the location 132km upstream (near Cross Creek waste water
treatment plant). Effect estimates of NH\(_3\), P and TKN consistently
hover near zero with two exceptions. The effect of \(NH_3\) appears to
decrease after 109km upstream, and the effect of TKN appears to increase
after 49km upstream.

The Supplementary Material includes summaries of results for all
cutpoints, space-varying confounding covariates, test statistic
distribution settings, as well as multiple outcome and exposure model
specifications. Point estimates varied modestly depending on the
cutpoint, space-varying confounding covariate, and how the
exposure/outcome models were specified. All of the point estimates for
NO\(_3\) were greater than zero for locations 109, 95, and 86 kilometers
upstream. Statistical significance was sensitive to the choice of Wald
test statistic distribution but generally accords with the shifts in
significance seen in the simulation results.

\section{Discussion}\label{discussion}

Our results corroborate existing evidence that the much more abundant
nitrate form of N is a major driver of downstream chlorophyll production
in the middle section of the Cape Fear River. Among dissolved inorganic
N species, cyanobacteria prefer to assimilate N in the form of ammonium
and then switch to nitrate uptake when ammonium is depleted
\citep{burkholder2002}. In this section of the Cape Fear River, ammonium
is typically at low concentrations while nitrate concentrations are an
order of magnitude higher \citep{mallin2006, kennedy2008}. Experimental
additions of inorganic and organic N have stimulated algae growth in the
Cape Fear River \citep{dubbs2008} and its two major tributaries, the
Black and Northeast Cape Fear Rivers \citep{mallin2004}. As
cyanobacteria are a primary harmful algal taxa group of concern in this
system, it is notable that N stimulates growth of this group
\citep{burkholder2002} as well as growth of \emph{Microcystis}
specifically \citep{paerl1987, siegel2011, yuan2014}.

Both point and nonpoint sources such as agricultural runoff contribute
to nutrient concentrations in the Cape Fear River
\citep{rajbhandari2015}. Our data cannot distinguish between sources of
pollution. While the data also cannot precisely pinpoint locations of
nutrient inputs into the river, our analysis does indicate areas for
further investigation. Across our choices of cutpoints and various
exposure and outcome models (see Supplementary Material for details),
significant effects of NO\(_3\) tended to occur between 86 and 109km
upstream of LD1. Notably, four major National Pollutant Discharge
Elimination System permits are located in this reach of the river: the
Tarheel Plant (NC0078344), Dupont Fayetteville Works (NC0003573), Cedar
Creek Site (NC0003719), and the Rockville Creek Waste Water Treatment
plant (NC0050105).

We have shown how ``what if'' questions on water quality of scientific
and policy interest can be mathematically framed as causal estimands. In
the presence of space- and/or time-varying confounding, this must be
accounted for in estimation, else biased estimates will result. Our
application is one of the first to use the g-methods
\citep{robins2009estimation} as part of an ecological causal assessment.
In fact, despite their general utility, structural nested models have
rarely been applied in practice \citep{vansteelandt2014}. We demonstrate
how they can be implemented and give details for deriving a closed form
estimator in the Supplementary Material.

Results from observational studies can always be skeptically reviewed,
but the potential outcomes framework forms a basis for constructive
critique. The causal assumptions described in Section
\ref{causal-assumptions} must be thoroughly vetted. Has all confounding
been accounted for? Are the parametric forms of our models correctly
specified, or reasonably so? Various criteria have been proposed to
assess fit of marginal structural models
\citep{platt2013information, baba2017criterion}, and methods such those
described in \citet{wang2006diagnosing} may be used to check sensitivity
of IPW estimators. As a diagnostic of the g-formula model fit,
\citet{westreich2012parametric} suggest estimating the outcome
distribution based on the ``natural course'' of exposure, which should
be comparable the observed outcome distribution. While developed in the
context of inferring optimal dynamic treatment regimes,
\citet{rich2010model} and \citet{henderson2010regret} proposed methods
for assessing fit based on residuals of the models within a structural
nested model.

As with any research, the causal effects estimated from a single
analysis should not be the sole source of evidence. The methods
described in this paper can augment assessment and decision frameworks
such as the EPA's CADDIS. The potential outcomes framework is well
suited to aid policymakers in development of permit standards and
surface water standards. \citet{yuan2010}, for example, estimated
effects of nutrient pollution on stream invertebrates using propensity
score methods.

Lastly, while GEE methods are available in several R packages
\citep{geepack, carey2012gee}, generic M-estimation is not
straightforward with current statistical software. The causal models
implemented in this paper involve combining estimating equations from
several models. The R package geex was developed in conjunction with
this research to streamline a programmer's work in implementing
estimating equation theory. The Supplementary Material includes the code
to replicate our analyses, and examples of geex for causal models may be
found therein.

\bigskip

\begin{center}
{\large\bf SUPPLEMENTARY MATERIAL}
\end{center}

\begin{description}

\item[Appendices:] Four appendices containing additional explication of the g-formula, derivation of the closed form structural nested model estimators, details on simulations parameterizations, and additional analyses of Cape Fear River data. (PDF file)
\item[R-package updown:] R-package updown containing code to perform the simulations and analyses described in this paper. (GNU zipped tar file)
\item[R-package geex:] R-package geex containing code necessary for sandwich variance estimators used in updown package. (GNU zipped tar file)
\item[R-package capefear:]  R-package capefear containing data from the Cape Fear River obtained by the North Carolina Nature Conservancy. Also contains discharge data from USGS stream gauges for the period of the study. (GNU zipped tar file)

\end{description}

\nocite{rubin2004multiple}

\bibliographystyle{chicago}

\bibliography{updownstream.bib}

\newpage 
 \vspace{3mm}
  \begin{center}
  \LARGE\bf Supplementary Material for \\
  \medskip
    {\LARGE\bf Upstream Causes of Downstream Effects}    
\end{center}
  \medskip

\appendix

\section{Parametric g-formula formulation}\label{gfm-derivation}

In all of the analyses, the outcome model was parameterized within the
g-formula as a linear model:

\begin{equation}
\tag{G1}\label{G1}
\operatorname{E}\left[ Y_{3t} | \overline{A}_{t} = [0_{t -1}:a_t], \overline{O}_t = \overline{o}_t \right] = h(\overline{A}, \overline{O}, \beta)
\end{equation}

\noindent For example, the correctly specified \(h\) for the simulations
is \[
\beta^{g}_{0} + \beta^{g}_1 a_{2t} + \beta^{g}_2 a_{1t} + \beta^{g}_3 l_{12t} + \beta^{g}_4 l_{22t} + \beta^{g}_5 y_{3, t - 1 }.
\]

\noindent The stability analyses varied which \(L\) covariates were
included in \(h\), but the parameterization of the exposure (\(a_{2t}\)
and \(a_{1t}\)) was not modified.

The average potential outcome
\(\operatorname{E}\left[ Y_{3t}([0_{t -1}:a_t]) \right]\) can be linked
to observed data in the following manner:

\begin{align*}
\tag{G2}\label{G2}
& \operatorname{E}\left[ Y_{3t}([0_{t -1}: a_t])  \right] \\
& = \operatorname{E}\left\{ \operatorname{E}\left[Y_{3t}([0_{t -1}:a_t]) | \overline{A}_{2t} = [0_{t -1}: a_t], \overline{L}_t = \overline{l}_t \right] \right\} \text{ (no unmeasured confounders) } \\
& = \operatorname{E}\left\{ \operatorname{E}\left[Y_{3t} | \overline{A}_{t} = [0_{t -1}: a_t], \overline{L}_t = \overline{l}_t \right] \right\} \text{ (causal consistency) } \\
\end{align*}

According to the parametric g-formula, models for each
\(\operatorname{f}_{l_{pjk}} = \operatorname{f}_{l_{pjk}| \overline{l}_{j-1, k-1}, \overline{a}_{j-1, k-1}}\)
must be fit. However, as will be clear below, the parameters
corresponding to non-space- or time-varying covariates cancel in a
causal contrast. Hence, we need only fit a model for the conditional
mean of \(L_{22t}\), for which we used a standard linear model with
expectation
\(\gamma^g_0 + \gamma^g_1 l_{12t} + \gamma^g_2 l_{22, t - 1} + \gamma^g_3 a_{1t}\).
By causal consistency,
\(\operatorname{E}[L_{22t}(a_{1t})] = \gamma^g_0 + \gamma^g_1 l_{12t} + \gamma^g_2 l_{22, t - 1} + \gamma^g_3 a_{1t}\).
Under this assumed parameterization,
\(\operatorname{E}[L_{22t}(a_{1t}) - L_{22t}(0)] = \gamma^g_3\).

Putting \eqref{G2} together with the model for \(f_{l_{22t}}\) obtains:

\begin{align*}
& \operatorname{E}\left[ Y_{3t}(0_{t -1}: a_t) - Y_{3t}(0_t) \right] \\
& = \operatorname{E}\left\{ \operatorname{E}\left[ Y_{3t}(0_{t -1}: a_t) - Y_{3t}(0_t) \right] \bigg| A_{1t} = 0, A_{2t} = 0_{t -1}: a_t, \overline{L}_{2t} \right\} \\
& = \operatorname{E}\left\{ \beta^{g}_1 a_{2t} + \beta^{g}_2 a_{1t} + \beta^{g}_4[L_{22t}(a_{1t}) -  L_{22t}(0) ]\right\} \text{ (plugging in $h$) }\\
& = \beta^{g}_1 a_{2t} + \beta^{g}_2 a_{1t} + \beta^{g}_4 \operatorname{E} \left\{ \operatorname{E} [L_{22t}(a_{1t}) -  L_{22t}(0) | A_{11} = a_{11}, L_{12t}, L_{22, t - 1} ] \right\} \\
& = \beta^{g}_1 a_{2t} + \beta^{g}_2 a_{1t} + \beta^{g}_4 \gamma^g_3. \\
\end{align*}

\noindent When \(a_t = (1, 1)'\) for all \(t\),
\(\mu^g = \frac{1}{n_t} \sum_{t = 1}^{n_t} \operatorname{E}\left[ Y_{3t}([0_{t -1}, a_t]) - Y_{3t}(0_t) \right] = \frac{1}{n_t} \sum_{t = 1}^{n_t} [ \beta^{g}_1 + \beta^{g}_2 + \beta^{g}_4 \gamma^g_3]\).

\section{Closed form estimator for SNM parameters}\label{snm-closed}

Vansteelandt and Joffe (2014) show that a consistent estimator of
\(\beta^{s}\) can found by solving estimating equations (eq. 33):

\[
\sum_i \sum_t \sum_s \{ \operatorname{d}_{st}(\overline{L}_{ist}, \overline{A}_{ist}) - \operatorname{E}[\operatorname{d}_{st}(\overline{L}_{ist}, \overline{A}_{ist}) | \overline{L}_{ist}, \overline{A}_{i,s - 1,t - 1}] \} \{ U_{ist}(\beta^{s}) - \operatorname{E}[U_{ist}(\beta^{s})| \overline{L}_{ist}, \overline{A}_{i, s - 1,t - 1}] \}
\]

\noindent where
\(\operatorname{d}_{st}(\overline{L}_{ist}, \overline{A}_{ist})\) is
chosen to be
\(\operatorname{E}[ \partial U_{st}(\beta^{s})/\partial \beta^{s} | \overline{L}_{ist}, \overline{A}_{ist}]\).
This formulation is slightly different from Vansteelandt and Joffe in
that we added an additional dimension \(s\). Since our endogenous
covariate is space-varying rather than time-varying, the blip process is
indexed by \(s\) rather than \(t\).

Let
\(\rho_{ist} = \operatorname{E}[\operatorname{d}_{st}(\overline{L}_{ist}, \overline{A}_{ist}) | \overline{L}_{ist}, \overline{A}_{i,s - 1,t - 1}]\)
and
\(\lambda_{ist} = \operatorname{E}[U_{ist}(\beta^{s})| \overline{L}_{ist}, \overline{A}_{i, s - 1,t - 1}]\),
then:

\begin{align*}
& \sum_i \sum_t \left\{ [\operatorname{d}_{1t}(\overline{L}_{i1t}, \overline{A}_{i1t}) - \rho_{i1t}] (U_{i1t}(\beta^{s}) - \lambda_{i1t}) + [\operatorname{d}_{2t}(\overline{L}_{i2t}, \overline{A}_{i2t}) - \rho_{i2t}] ( U_{i2t}(\beta^{s}) - \lambda_{i2t} ) \right\} \\
&= \sum_i \sum_t \left\{ \left[ \begin{pmatrix} A_{i2t} \\ 0 \end{pmatrix} - \rho_{i2t} \right] (U_{i1t}(\beta^{s}) - \lambda_{i1t}) + \left[ \begin{pmatrix} A_{i2t} \\ A_{i2t} \end{pmatrix} - \rho_{i2t} \right] ( U_{i2t}(\beta^{s}) - \lambda_{i2t})  \right\} \\
&= \sum_i \sum_t \left\{ \left[ \begin{pmatrix} A_{i2t} \\ 0 \end{pmatrix} - \rho_{i2t} \right] (Y_{i3t} - \beta^{s}_1 A_{i2t} - \lambda_{i1t}) \right. \\
& \quad \quad \quad \left. + \left[ \begin{pmatrix} A_{i2t} \\ A_{i2t} \end{pmatrix} - \rho_{i2t} \right] ( Y_{i3t} - \beta^{s}_1 A_{i2t} - \beta^{s}_2 A_{i1t} - \lambda_{i2t} )  \right\} \\
&= \sum_i \sum_t \left\{  \begin{pmatrix} B^0_{i2t} (r_{i3t} - \beta^{s}_1 A_{i2t} ) +  B^1_{i2t} (r^0_i - \beta^{s}_1 A_{i2t} - \beta^{s}_2 A_{i1t} ) \\ B^1_{1it} (r^1_i - \beta^{s}_1 A_{i2t} - \beta^{s}_2 A_{i1t} )  \end{pmatrix} \right\}. \\
\end{align*}

\noindent In the last line, we let \(B_{ist} = A_{i2t} - \rho^k_{ist}\).
Let \(r_{ist} = Y_{i3t} - \lambda^k_{ist}\).

Let \(C_i = \sum_t r^0_i (B^0_{i2t} + B^1_{i2t})\),
\(D_i = \sum_t (B^0_{i2t} A_{i2t} + B^1_{i2t} A_{i1t})\),
\(E_i = \sum_t B^1_{i2t} A_{i1t}\), \(F_i = \sum_t B^1_{1t} r^1_i\),
\(G_i = B^1_{i1t} A_{i2t}\), and \(H_i =\sum_t B^1_{i1t} A_{i1t}\). Then
\(\beta^{s}\) is the solution to:

\begin{align*}
\sum_i \begin{pmatrix}
C_i - \beta^{s}_1 D_i - \beta^{s}_2 E_i \\
F_i - \beta^{s}_1 G_i - \beta^{s}_2 H_i
\end{pmatrix}
\end{align*}

\noindent which yields,

\[
\hat{\beta}^{s}_1 = \frac{\sum_i E_i \sum_i F_i  - \sum_i C_i \sum_i H_i }{\sum_i E_i \sum_i G_i - \sum_i D_i \sum_i H_i }  \text{ and } \hat{\beta}^{s}_2 = \frac{\sum_i D_i \sum_i F_i  - \sum_i C_i \sum_i G_i }{\sum_i D_i \sum_i H_i - \sum_i E_i \sum_i G_i } 
\]

\noindent where

\[
\sum_i E_i \sum_i G_i \neq \sum_i D_i \sum_i H_i , \quad \sum_i E_i \neq 0.
\]

\section{Simulation details}\label{simparms}

The nodes in the simulated study system were parameterized and generated
according to the following distributions:

\begin{align*}
\label{simparms}
t = 0 & \begin{cases}
L_{110} & \sim N(21.5, 2.5) \\
L_{220} & \sim N(-2.8, 0.7) \\
A_{s0} & \sim Bern(0.1) \text{ for $s = 1, 2$} \\
Y_{30} & \sim N(2.25, 1.25) 
\end{cases} \\
t = 1, 2, 3 & \begin{cases}
L_{11t} & \sim N(23 + 0.2 L_{11, t - 1}, 2) \\
A_{1t} & \sim Bern(\mbox{logit}^{-1}(-2.5 + 0.09 L_{11t} + 0.025 A_{1, t - 1})) \\
L_{12t} & \sim N(6.75 + 0.75 L_{11, t - 1}, 1) \\
L_{22t} & \sim N(2 - 0.04 L_{12t} + 0.04 L_{22, t - 1} + 0.3 A_{1t}, 0.25) \\
A_{2t} & \sim Bern(\mbox{logit}^{-1}(-2.5 + 0.09 L_{12t} + 0.1 L_{22t} +  0.05 A_{1t} + 0.025A_{2, t- 1})) \\
Y_{3t} & \sim N(-5 + 1 A_{s - 1, t} + 0.5 A_{s - 2, t}  + 0.025 L_{1,s - 1, t} + 0.5 L_{2,s - 1, t} + 0.35 Y_{3, t - 1}, 1) 
\end{cases}
\end{align*}

\noindent Code for the simulations can be found in the updown R package
of the Supplementary Materials.

\section{Stability analyses}\label{cfrdetails}

In addition to estimating the target parameters using all possible
combinations of settings of the cutpoint and space-varying confounding
variables, we also modified the exposure and outcome models to include a
temperature by flow interaction term in both outcome and exposure
models. This resulted in a total of 1200 point estimates per method. If
some component model failed to converge for a method, then the estimate
attempt was considered a failure. Across all 3600 attempts, model
fitting failed 272 times for the MSMs, 70 times for the SNMs, and zero
times for the g-formula.

To check the stability for the primary results presented in Figure 5 to
the simple imputation procedure described in the main text, a multiple
imputation procedure was used. The \texttt{aregImpmute} function from
the Hmisc R package (Harrell et al., 2017) was used to impute missing
values using year, month, distance between sites, flow, temperature, pH,
dissolved oxygen, and each of the NP species. Five complete datasets
were generated using the predictive mean matching imputation method with
the match=`closest' option. Point estimates from the complete datasets
were averaged and the corresponding estimated standard errors pooled
using Rubin's rules (Rubin, 2004).

\begin{figure}
\centering
\caption{This figure shows the 3258 point estimates  obtained across settings of the cutpoint, space-varying confounding variables, and exposure and outcome models. Point estimates for NO$_3$ between 86 and 109 km upstream are positive across all settings and have the most consistently strong effects. TKN estimates for the two locations just upstream of LD1 are all positive, while the P estimates from the same locations are all negative.}
\label{fig:cfrestimates}
\includegraphics{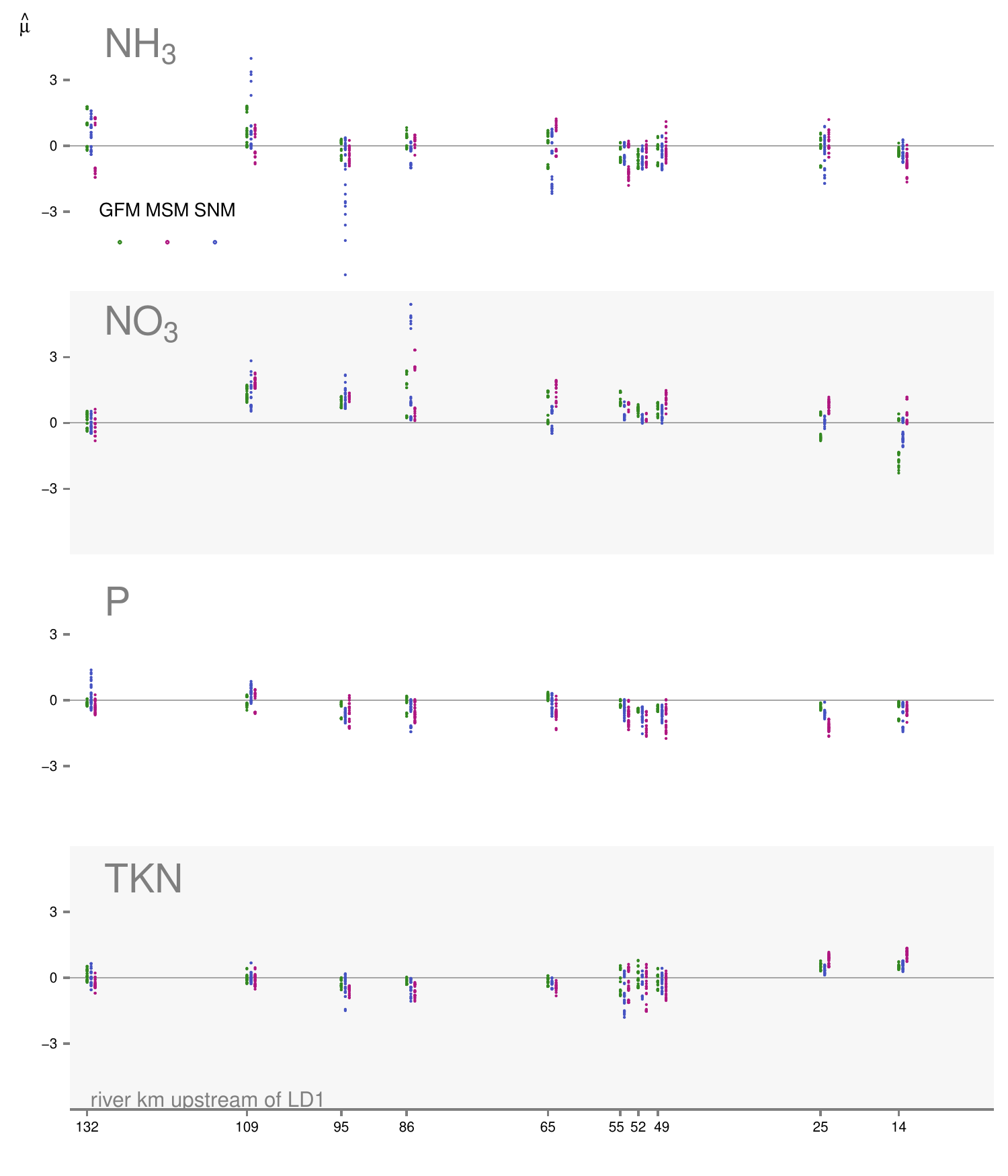}
\end{figure}

\begin{figure}
\centering
\caption{Volcano plot of estimates of the Cape Fear analysis showing significance and estimates for all analysis methods and models. The point estimates are same in all the panels. Significance levels vary depending on the distribution used for the test statistic.}
\label{fig:cfrvolcano}
\includegraphics{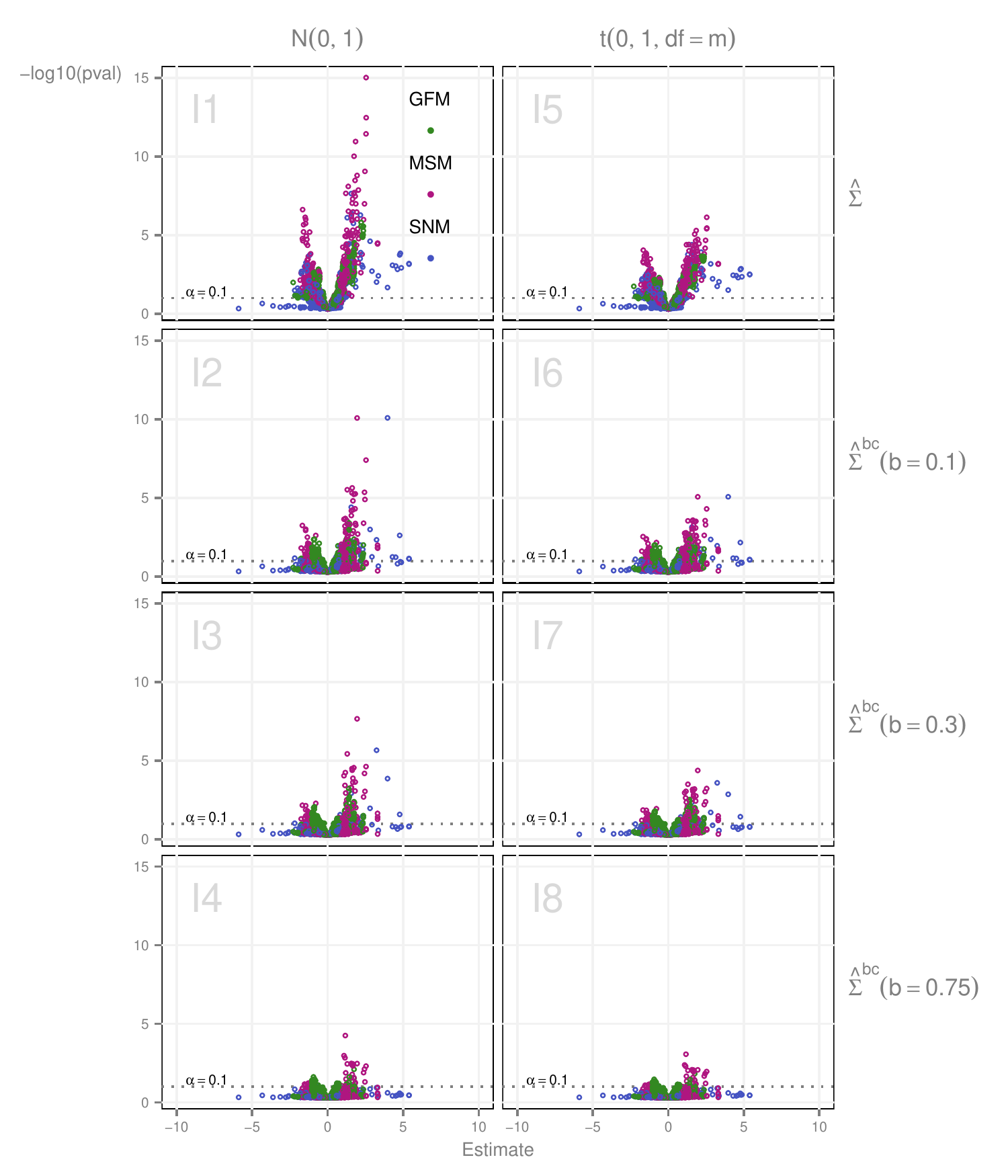}
\end{figure}

\begin{figure}
\centering
\caption{This figure repeats main text Figure 5 with estimates based on the estimated derived from 5 imputed datasets. These results are largely substantively the same as those reported in main text Figure 5.}
\label{fig:cfrimputation}
\includegraphics{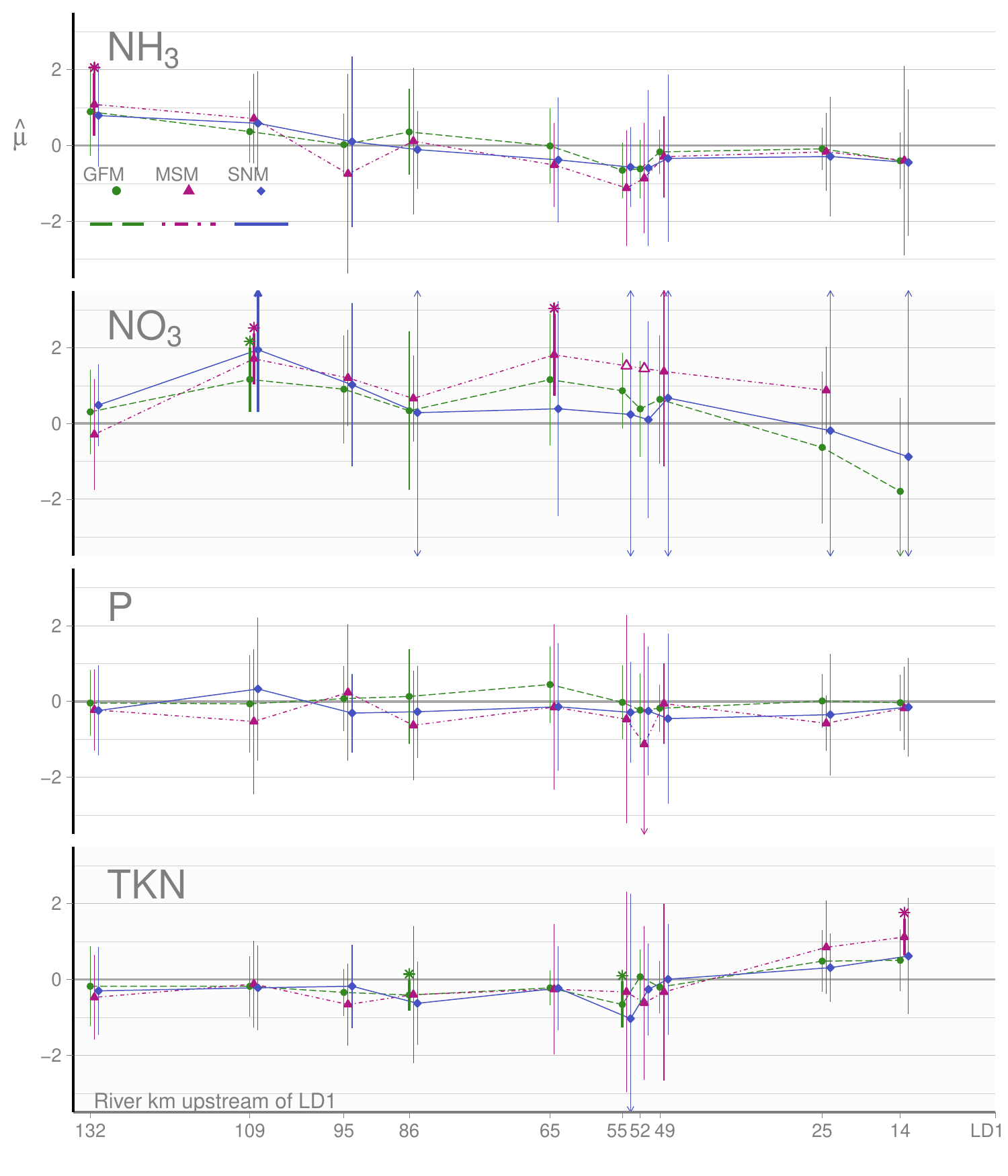}
\end{figure}

Figure \ref{fig:cfrestimates} shows all the point estimates within (-6,
6) across all the model options. In two cases, estimates from an SNM
were outside this range. The results in Figure \ref{fig:cfrestimates}
conform to the general patterns described in the main text. Figure
\ref{fig:cfrvolcano} show all the point estimates along with the
p-values using different test statistic distributions. In each panel,
the point estimates are the same, but the significance clearly depends
on the test statistic distribution. Figure \ref{fig:cfrimputation}
repeats Figure 5 from the main text using estimates based on the
imputation procedure described above.

\bibliographystyle{plainnat}

\end{document}